\documentclass[sigconf,screen]{acmart}

\usepackage{booktabs}
\usepackage{multirow} 
\usepackage{pifont}
\usepackage{graphicx}
\usepackage{makecell}
\usepackage{adjustbox}
\usepackage{forest}
\usepackage{subcaption}
\usepackage{wrapfig}
\usepackage{tablefootnote}
\usepackage{threeparttable}
\usepackage{xspace}
\usepackage{algorithm}
\usepackage{algpseudocode}
\usepackage{algorithmicx}

\usepackage{colortbl}
\usepackage{color}
\usepackage{xcolor}
\usepackage{makecell}
\usepackage{enumitem}
\usepackage{threeparttable}

\definecolor{tabletitle}{HTML}{E5DBF8} 
\definecolor{lightblue}{HTML}{DAE8FC}
\colorlet{highlight}{lightblue!60}

\definecolor{checkgreen}{HTML}{009900}
\definecolor{crossred}{HTML}{CC0000}

\newcommand{\cmark}{\textcolor{checkgreen}{\ding{51}}}
\newcommand{\xmark}{\textcolor{crossred}{\ding{55}}}
\newcommand{\up}[1]{\textcolor{crossred}{\,\textuparrow #1\%}}
\newcommand{\down}[1]{\textcolor{checkgreen}{\,\textdownarrow #1\%}}
\newcommand{\fast}[1]{\textcolor{crossred}{\,#1$\times$}}
\newcommand{\nofast}[1]{\textcolor{checkgreen}{\,#1$\times$}}

\usepackage{mdframed}
\newcounter{rq} 
\newcommand{\answerRQ}[1]{\refstepcounter{rq}
\vspace{1mm}
\begin{mdframed}[linecolor=gray,roundcorner=12pt,backgroundcolor=gray!15,linewidth=3pt,innerleftmargin=2pt, 
leftmargin=0cm, rightmargin=0cm, topline=false, bottomline=false, rightline=false]
\textbf{RQ\arabic{rq} Summary:} #1
\end{mdframed}
\vspace{1mm}
}

\newcommand{\appname}{\textsc{DyCoder}\xspace}

\newcommand{\retriever}{\textsc{DyRetriever}\xspace}

\AtBeginDocument{%
  }

\usepackage{microtype}
\setcopyright{cc}
\setcctype{by}
\acmDOI{10.1145/3832783.3834350}
\acmYear{2026}
\copyrightyear{2026}
\acmISBN{979-8-4007-2882-2/2026/10}
\acmConference[ASE '26]{Proceedings of the 41st IEEE/ACM International Conference on Automated Software Engineering}{October 12--16, 2026}{Munich, Germany}
\acmBooktitle{Proceedings of the 41st IEEE/ACM International Conference on Automated Software Engineering (ASE '26), October 12--16, 2026, Munich, Germany}
\acmSubmissionID{ase26main-p356-p}
\received{2026-03-26}
\received[accepted]{2026-06-18}

\begin{document}

\title{Effective and Efficient Context Retrieval via Partial Dependency Graph for Repository-Level Code Generation}

\author{Zhongxin Liu}
\email{liu\_zx@zju.edu.cn}
\affiliation{%
  \institution{College of Computer Science and Technology and the State Key Laboratory of Blockchain and Data Security, Zhejiang University}
  \city{Hangzhou}
  \country{China}
}
\author{Zhonghao Jiang}
\email{zhonghao.j@zju.edu.cn}
\affiliation{%
  \institution{College of Computer Science and Technology and the State Key Laboratory of Blockchain and Data Security, Zhejiang University}
  \city{Hangzhou}
  \country{China}
}
\author{Zhifan Ye}
\email{yezhifan@zju.edu.cn}
\affiliation{%
  \institution{College of Computer Science and Technology and the State Key Laboratory of Blockchain and Data Security, Zhejiang University}
  \city{Hangzhou}
  \country{China}
}
\author{Haoye Wang}
\authornote{Corresponding author.}
\email{wanghaoye@hzcu.edu.cn}
\affiliation{%
  \institution{School of Computer and Computing Science, Hangzhou City University}
  \city{Hangzhou}
  \country{China}
}
\author{Jiakun Liu}
\email{jiakunliu@hit.edu.cn}
\affiliation{%
  \institution{Faculty of Computing, Harbin Institute of Technology}
  \city{Harbin}
  \country{China}
}
\author{Xiaoxue Ren}
\email{xxren@zju.edu.cn}
\affiliation{%
  \institution{School of Software Technology and the State Key Laboratory of Blockchain and Data Security, Zhejiang University}
  \city{Hangzhou}
  \country{China}
}


\renewcommand{\shortauthors}{Zhongxin Liu, Zhonghao Jiang, Zhifan Ye, Haoye Wang, Jiakun Liu, Xiaoxue Ren}

\begin{abstract}

LLM-based repository-level code generation aims to generate code according to the available context within a software repository, requiring LLMs to understand and reason over complex code dependencies. 
Due to limited context windows and insufficient understanding of repository-specific code, LLMs typically rely on retrieval-augmented generation (RAG) to incorporate relevant contextual code. 
Early RAG approaches primarily employ similarity-based retrieval, which often fails to retrieve code snippets that the target function actually depends on.
Recent work introduces graph-based retrieval to model such dependencies, but typically relies on manually designed rules and static global graphs, leading to limited flexibility and high construction and maintenance costs.

In contrast, human developers typically collect helpful context by implicitly constructing a partial dependency graph and iteratively inspecting along it.
Inspired by human behavior, we propose \retriever, an effective and efficient context retrieval method via partial dependency graphs.
\retriever leverages an LLM to simulate how human developers collect helpful context by first selecting a set of entry-point functions and then performing multi-hop reasoning along the code dependency graph.
During multi-hop reasoning, \retriever leverages the LLM’s semantic understanding to explicitly validate whether a function can help to generate the target function, thereby eliminating reliance on manually designed rules and enabling flexibility across application scenarios.
Instead of statically constructing a global dependency graph, \retriever builds a partial graph on demand and discards it after use, reducing construction overhead and avoiding maintenance costs.
We integrate \retriever with a similarity-based code retriever to build a new code generation approach, \appname, and evaluate it on two widely used repository-level code generation benchmarks, CoderEval and DevEval.
Experimental results show that \appname achieves relative Pass@1 improvements of 25.63\% and 59.73\% on CoderEval and DevEval, respectively, compared with existing RAG-based methods, while being 7.4$\times$ faster than baselines based on static dependency graph construction.

\end{abstract}

\maketitle

\section{Introduction}

Code generation aims to automatically generate the code implementation corresponding to a given specification~\cite{wang2023review, jiang2024survey, tao2025retrieval, sun2026fly, sun2026cost}.
Large language models (LLMs)~\cite{deepseekv3.2, qwen3, hurst2024gpt}, owing to their strong capabilities in language understanding and generation, have demonstrated remarkable performance in code generation~\cite{evalplus, bigcodebench, livecodebench, SWE-bench, pan2025modularization}.
Recently, repository-level code generation, which aims to generate a target function according to its signature\footnote{In this paper, we consider a function’s signature to include its corresponding docstring.} and available context in the repository, has attracted widespread attention due to its close alignment with real-world development scenarios~\cite{jiang2025agentic, tao2025retrieval}.
However, due to their limited context window and lack of understanding of specific code repositories, it is difficult for LLMs to handle complex repository-level code generation tasks.
To mitigate this challenge, recent studies~\cite{aligncoder, rlcoder, reposcope, repocoder} typically adopt the retrieval-augmented generation (RAG) paradigm, wherein the signature of the target function is used as queries to retrieve helpful code snippets from the repository, and the retrieved code is then provided to the LLM as the context for code generation.

The key challenge for RAG-based methods is retrieving contexts that accurately guide the LLM to generate correct target functions within a repository.
Prior works~\cite{repocoder, rlcoder, aligncoder} mainly retrieve similar code snippets through sparse retrieval (e.g., BM25~\cite{BM25, lu2022reacc}) or dense retrieval (e.g., RepoCoder~\cite{repocoder}) as the contexts, which have been shown to be beneficial.
However, similarity-based retrieval is often insufficient for generating the target function~\cite{rlcoder, graphcoder, draco}, since it may overlook code snippets that the target function is not similar to, but actually depends on.
Therefore, advanced methods incorporate graph-based retrieval to supplement similarity-based retrieval to capture semantic relationships, including dependencies~\cite{graphcoder} and containment~\cite{reposcope}, that similarity alone cannot reveal.
Specifically, these methods construct code graphs, such as program dependency graphs~\cite{reposcope, graphcoder}, of the whole repository using static analysis tools, and design retrieval algorithms using manually crafted rules to perform retrieval based on reachability on code graphs.

However, \emph{such designs often lack flexibility in capturing the diverse and complex dependency patterns that arise in real-world repositories.}
Existing studies have~\cite{liao20243, guan2024contextmodule} demonstrated that repository-level code generation in different application scenarios requires different types of context, e.g., in-file code snippets, cross-file dependencies, API, or symbol definitions.
Nevertheless, existing graph-based methods typically rely on fixed rules that are tailored to a specific type of context, such as handcrafted similarity metrics~\cite{graphcoder} and import-based traversal~\cite{reposcope}, making them difficult to adapt when the required contextual types change across tasks.
For example, RepoScope~\cite{reposcope} retrieves cross-file context by traversing the code graph from entities imported by the target function, and thus cannot obtain such context when relevant cross-file dependencies are not explicitly imported.
Moreover, \emph{the complete dependency graph of the whole repository is both time-consuming to parse and difficult to maintain}.
For example, it may take several hours to construct the dependency graph for a large codebase with more than 100,000 lines of code~\cite{fan2020escaping, keshani2024frankenstein}. 
Such overhead is often unacceptable for practical code generation, as developers typically expect to obtain code within a short time frame~\cite{tao2025retrieval, yang2025empirical, li2025reasoning}.
In addition, developers continuously modify and extend code within a repository. 
As development progresses, dependency relations between code elements also evolve, requiring frequent reconstruction or updating of the dependency graph and thus incurring additional maintenance costs.

In contrast to existing graph-based retrieval methods, human developers typically do not rely on predefined retrieval rules to locate relevant context.
Instead, they flexibly inspect code based on their understanding of the task and the repository, iteratively judging whether a function is helpful for completing the target function~\cite{fleming2013information}.
Moreover, they do not pre-construct or maintain a complete repository-level dependency graph.
They usually start from a small number of functions that appear relevant as entry points and explore related functions on demand by following dependency chains~\cite{lawrance2010programmers, latoza2010developers}.
Through this process, developers implicitly construct a partial dependency graph during inspection, rather than relying on a global one~\cite{Investigating2025}.

Inspired by how developers iteratively collect helpful contexts, we propose \retriever, an effective and efficient context retrieval method via partial dependency graphs. 
\retriever retrieves functions that are useful for generating the target function by simulating the behavior of human developers with the help of an LLM.
It first leverages an LLM to identify some functions as entry points based on the signature of the target function and the repository structure, mimicking human developers' initial selection.
Starting from these entry points, \retriever explores related functions through multi-hop reasoning, reflecting how developers progressively decide whether additional dependencies are helpful.
However, such decisions are inherently dependent on developers’ understanding of a function and its relationship to the target function, which are hard to adequately model with pre-defined rules.
We therefore leverage an LLM, whose strong semantic understanding of code enables it to determine at each hop whether exploring additional dependencies is likely to be helpful.
In addition, to support such multi-hop reasoning, \retriever requires local dependency information that can be obtained on demand.
However, existing static analysis tools focus on providing global dependency graphs and do not support efficient construction of partial dependency graphs~\cite {salis2021pycg, bouzenia2024dypybench}.
Considering the callees of a function can usually be inferred from the code and its imported modules, we further employ the LLM to dynamically construct partial dependency graphs, enabling human-like on-demand construction of a partial dependency graph during retrieval.

\begin{figure*}
    \centering
    \includegraphics[width=1\linewidth]{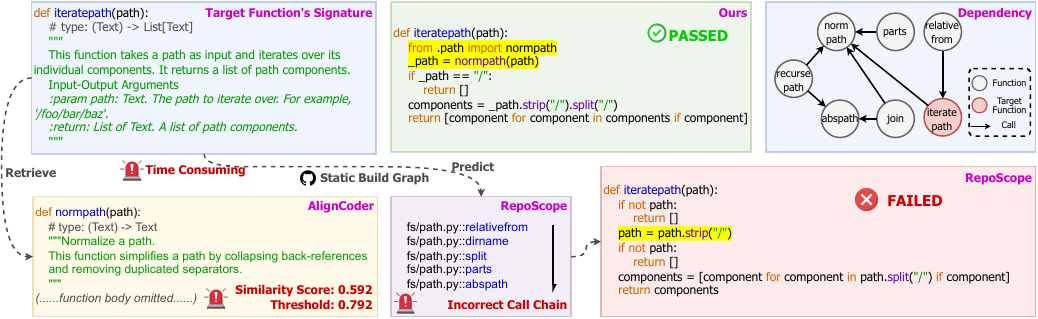}
    \caption{A motivating example of \textit{``fs.path.iteratepath''} in DevEval.}
    \label{fig:motivation}
\end{figure*}

\retriever can supplement similarity-based retrieval by leveraging semantic information that is not captured by similarity.
By integrating \retriever with similarity-based code retrieval, we further propose a new repository-level code generation approach named \appname.
We evaluate \appname on two popular repository-level code generation benchmarks, CoderEval~\cite{CoderEval} and DevEval~\cite{DevEval}, with three advanced LLMs, and perform comprehensive comparisons with state-of-the-art baselines.
Experimental results show that, compared with existing RAG methods, \appname achieves an average relative improvement of 25.63\% and 59.73\% in Pass@1 on CoderEval and DevEval, respectively.
The results of our ablation studies show that removing either the \retriever or the similarity-based retrieval leads to a comparable average performance drop of 14.80\% and 15.85\%, respectively, while removing multi-hop reasoning results in an average 6.55\% performance degradation. 
This demonstrates the complementarity between the two types of retrieval methods and reveals the effectiveness of multi-hop reasoning.
The results of our efficiency evaluation show that, compared with the static graph construction-based method, RepoScope, \appname achieves a 7.4 times improvement in efficiency without training cost, while attaining efficiency comparable to similarity-based methods.
Finally, we combine the context collected by \retriever with different similarity-based methods to verify the generalizability of \retriever.
The results show that \retriever can achieve 7.5\%-30.99\% performance improvement over existing similarity-based methods. 
This not only demonstrates the general applicability of \retriever, but also provides directions for future enhancements.

In summary, our main contributions are as follows:
\begin{itemize}[left=0pt, topsep=0em]
    \item We propose \retriever, a context retrieval method for repository-level code generation, inspired by human behaviors in collecting helpful code snippets. 
    \retriever uses LLMs’ semantic understanding to assess code snippets and dynamically construct partial dependency graphs, achieving good effectiveness and efficiency.
    \item We construct a new repository-level code generation approach, \appname, by integrating \retriever with a similarity-based retrieval method, and also demonstrate the generalizability of \retriever across different similarity-based retrieval methods. 
    \item Extensive and comprehensive experiments on CoderEval~\cite{CoderEval} and DevEval~\cite{DevEval} show that \appname achieves relative improvements of 25.63\% and 59.73\% in Pass@1, respectively, compared with other RAG methods, and is 7.4$\times$ faster than the baseline based on static graph construction.  
\end{itemize}

\section{Motivation}

A key challenge of repository-level code generation lies in extracting the contexts, which are helpful for the LLM to generate the target function, from the repository.
Existing methods typically retrieve code snippets based on text similarity or manually designed retrieval rules on the code graphs.
Figure~\ref{fig:motivation} presents a motivating example collected from the \texttt{PyFilesystem}~\cite{pyfilesystem2_github} repository.
This task aims to implement the \texttt{iteratepath} function, which returns a list of path components and depends on the \texttt{normpath} function defined in \textit{fs/path.py}. 
However, neither the state-of-the-art similarity-based method AlignCoder nor the state-of-the-art graph-based method RepoScope successfully retrieves \texttt{normpath} as context.

We further investigate why AlignCoder and RepoScope fail in this case.
The cosine similarity between the embeddings of the target function's signature and the function \texttt{normpath} is 0.592. 
AlignCoder only retrieves the top-5 most similar code snippets as context, leading to the similarity threshold for this case being 0.792.
Thus, AlignCoder fails to retrieve \texttt{normpath}.
This indicates that relying solely on similarity-based retrieval may overlook useful dependencies.
On the other hand, RepoScope designs an import-based retrieval rule such that only nodes reachable from entities imported by the target function within a certain distance in the dependency graph can be retrieved.
In this case, \texttt{normpath} is defined in the same file as \texttt{iteratepath} and does not meet the rule mentioned above.
Thus, RepoScope also fails to retrieve \texttt{normpath}.
Although this issue could be mitigated by adding extra heuristics (e.g., retrieving in-file functions), such case-by-case rules further highlight the limited flexibility of manually designed rules.

In practice, human developers can easily locate \texttt{normpath}.
For example, one may initially select \texttt{parts} and \texttt{join} based on their understanding of this repository since they may share partially similar program behaviors (as illustrated by the dependency graph in the top-right of Figure~\ref{fig:motivation}). 
When examining \texttt{parts} and \texttt{join}, the developer can trace their callees and find that \texttt{normpath} is a basic operation when processing paths and is used frequently by both \texttt{parts} and \texttt{join}.
Thus, the developer can successfully identify \texttt{normpath} as a relevant dependency for implementing \texttt{iteratepath}.
This observation suggests that human developers identify helpful context by first selecting some potential relevant functions as entry points and then iteratively following their dependency relations to explore additional relevant functions.
Since it is widely believed that LLMs operate similarly to humans, we may be able to collect context by simulating human developers using LLMs.

Achieving such a process requires constructing a dependency graph for the repository.
Existing works use static analysis tools to parse dependencies across the entire repository.
For example, RepoScope uses \texttt{PyType}~\cite{google_pytype} and \texttt{Tree-Sitter}~\cite{py_tree_sitter_docs} to parse relations such as calls, imports, and inherits to construct a static global graph.
However, statically constructing the global dependency graph of a repository is time-consuming~\cite{fan2020escaping, keshani2024frankenstein}.
For instance, RepoScope takes 261.61s to build the graph corresponding to this case.
In contrast, humans dynamically explore relevant parts by expanding a partial dependency graph. 
Inspired by the way humans analyze code, we propose \retriever, an effective and efficient context retrieval method for repository-level code generation.
It mimics human code analysis by leveraging the semantic understanding capabilities of LLMs and constructing and expanding partial dependency graphs.

\section{Approach}

\begin{figure*}[t!]
    \centering
    \includegraphics[width=\linewidth]{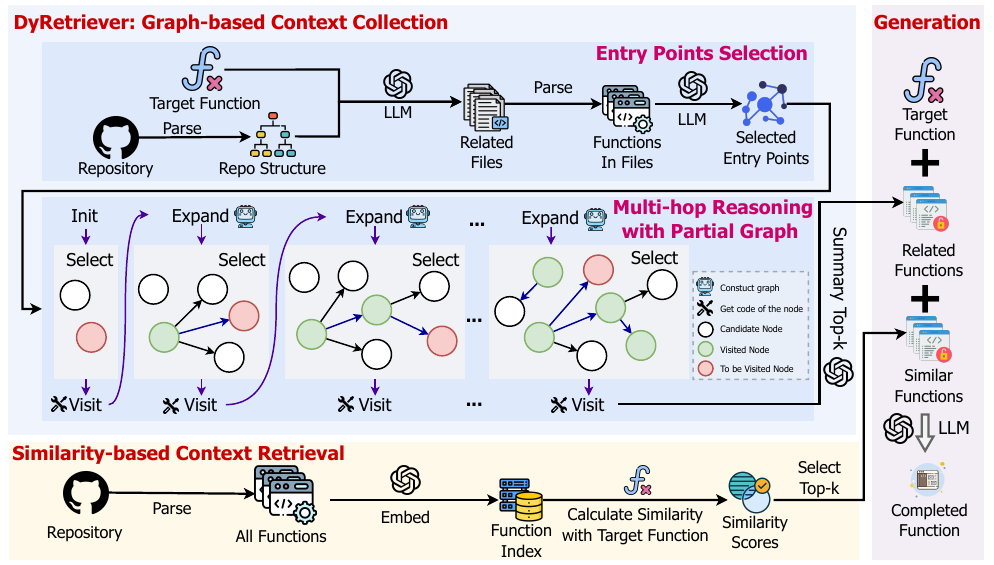}
    \caption{Overview of \appname.}
    \label{overview}
\end{figure*}

In this section, we introduce \appname, an LLM-driven repository-level code generation framework. 
Different from existing graph-based methods, \appname is a rule-free, autonomous framework that flexibly collects context through partial graphs.
As shown in Figure~\ref{overview}, \appname takes the signature of a target function and a repository as inputs and returns the completed code of the target function. 
It consists of three phases, i.e., graph-based context collection, similarity-based context retrieval, and generation.
In the graph-based context collection phase (Section~\ref{step1}), we introduce \retriever to perform effective and efficient retrieval. 
\retriever simulates the way human developers explore a repository when collecting contexts by reasoning over function dependencies.
Specifically, \retriever first initializes one or more functions as the entry points of the reasoning chain based on the provided repository structure. 
It then dynamically expands the partial code dependency graph and collects related functions as context according to the LLM’s traversal path over the graph nodes, which avoids the need to parse and store a global repository-level dependency graph, thereby significantly reducing time and maintenance overhead.
In the similarity-based context retrieval phase (Section~\ref{step2}), \appname first extracts all functions in the repository, and then computes the cosine similarity of embeddings generated by a pre-trained embedding model between each function and the signature of the target function to retrieve similar functions as context.
In the generation phase (Section~\ref{step3}), \appname deduplicates the context retrieved in the previous two phases, and then generates the code of the target function based on its signature and the deduplicated functions through prompt engineering.

\subsection{\retriever: Graph-based Context Collection}\label{step1}
This phase takes the signature of a target function and the repository as input and returns a list of functions in the repository.
Specifically, it consists of two stages, i.e., entry points selection and multi-hop reasoning with partial graphs.
The entry points selection stage (Section~\ref{step1.1}) aims to identify several functions from the code repository as entry points for multi-hop reasoning and partial graph construction.
It simulates how human developers, based on their understanding of the repository, identify several potentially relevant code snippets.
The multi-hop reasoning with partial graphs stage (Section~\ref{step1.2}) aims to iteratively perform graph expansion, node selection, and node visiting based on the entry point.
It simulates how human developers traverse dependency graphs and iteratively assess whether each function contributes to implementing the target function.

\subsubsection{Entry Points Selection}\label{step1.1}
This stage aims to establish the LLM’s understanding of the code repository and, based on the understanding, select several functions that may be relevant to the target function.
It reduces the search scope from the whole repository to several files that may be related to the target function.
It takes the signature of a target function and the repository as input and returns a list of functions as entry points.
To establish the LLM's understanding of the code repository, we first parse the repository into a \texttt{tree-based structure} following prior works~\cite{jiang2025agentic} to prevent exceeding the context window when letting the LLM read the entire repository.
The \texttt{tree-based structure} presents the organizational structure of files in the repository according to the folder containment hierarchy.
Then, to identify the entry points, we proceed from the file level to the function level to identify the entry points.
Specifically, based on the \texttt{tree-based structure} and the signature of the target function, the LLM is prompted to examine the overall layout of the codebase and return a list of file paths that are related to the target function, denoted as the related file list. 
To ensure that the subsequent reasoning process includes both in-file context and cross-file context, we additionally verify whether the LLM includes the file containing the target function.
If not, we explicitly add it to the related file list as the source of in-file context, while the remaining files serve as the source of cross-file context.
Next, we use a parser to extract all functions contained in the files from the related file list and construct a mapping, where the file path is used as the key, and all the function names contained in that file are used as the value.
Finally, we provide the signature of the target function and the mapping as context to the LLM, asking it to identify several function names as entry points.

\subsubsection{Multi-hop Reasoning with Partial Graphs}\label{step1.2}
This stage aims to perform multi-hop reasoning along the dependency graph based on the entry points and examine, one by one, whether the functions on the traversal path contribute to generating the target function.
Additionally, it dynamically constructs a function-level partial dependency graph instead of using a static parser to construct the entire graph of the repository to improve efficiency.
It takes the signature of the target function and the entry points as input and finally returns the top-k related functions.

As shown in Algorithm~\ref{AL1}, the process begins by initializing an empty list to record the traversal trajectory (Line 1) and a set of candidate functions using the entry points (Line 2). 
The \textit{candidate functions} form a global search space that is dynamically updated during exploration. 
We also set a maximum number of reasoning steps (Line 3), which serves as a global exploration budget.
The traversal algorithm consists of three steps: select, visit, and expand.

\begin{itemize}[left=0pt, topsep=0em]

\item \textbf{Select Step:} Select one function from the \textit{candidate functions} using the LLM (Line 5). The LLM is prompted to leverage the current context (i.e., previously selected functions and the target function signature) to identify a candidate function that best aligns with the target function.
\item \textbf{Visit Step:} Based on the function selected in the previous step, we determine whether this function has been visited before (Line 6); if so, this visit is skipped (Line 7), otherwise we obtain its source code (Line 9), update its visit status (Line 10-11), and record the visit trajectory in order (Line 12).
\item \textbf{Expand Step:} According to the function selected in the select step, all import statements from the code file in which the function resides are extracted (Line 14). 
Then, using the source code of this function obtained in the visit step and the import statements as input, the LLM is prompted to identify the functions it depends on (Line 15). 
To reduce hallucination during LLM identification, \retriever only requires the LLM to recognize two types of relations: call and lazy import.
Because import statements provide information about the source of APIs, the LLM is able to determine whether call or lazy import relations are valid based on the code. 
This step expands the partial dependency graph dynamically and retrieves the neighbor nodes of the currently visited function. 
\retriever subsequently adds the functions identified by the LLM to the \textit{candidate functions} to enable expansion before the next traversal iteration (Line 16).

\end{itemize}

After performing a limited number of traversal iterations, the multi-hop reasoning algorithm uses the trajectory collected during traversal as context and prompts the LLM, based on the target function, to determine the most related top-k functions from the trajectory (Line 20-21) as the semantic-based context.

While the LLM can efficiently identify related functions from the traversal trajectory, we observe two major issues when directly prompting it to expand the dependency graph.
First, hallucinations of the LLM may fabricate non-existent functions.
Second, APIs from standard libraries may be mistakenly reported as cross-file APIs. 
For example, the \texttt{match} API from the \texttt{re} library may be incorrectly identified by the LLM as one of the function’s dependencies.
Since we focus only on cross-file API dependencies, such dependencies should be discarded.
To address these two issues, we post-process the dependency graph constructed by the LLM. 
Specifically, for each dependency function identified by the LLM, \retriever searches the code repository for its corresponding implementation by matching the function name.
If no implementation is found, the dependency is discarded.
This step improves the reliability of the constructed dependency graph and enhances the effectiveness of \retriever.

\begin{algorithm}
\caption{Multi-hop Reasoning with Partial Graphs}
\begin{algorithmic}[1]
    \Statex \textbf{Input:} tarFunc \Comment{Signature of the target function}
    \Statex \textbf{Input:} entries \Comment{List of entry point functions including file names and function names}
    \Statex \textbf{Input:} maxHop \Comment{Maximum steps of multi-hop reasoning}
    \Statex \textbf{Output:} contextFunctions \Comment{Functions retrieved by \retriever as contexts}
    
    \State trajs = [] \Comment{Record traversal path}
    \State candidates = entries \Comment{To be visited}
    \State hop = 0
    \While{hop$<$maxHop} \Comment{Start Traversal}
    \State selectedFunc = $\text{LLM}_{select}$(tarFunc, candidates, trajs) \Comment{Select step}
    \If{isVisited(selectedFunc)}
        \State \textbf{continue}
    \EndIf
    \State funcContent = getCode(selectedFunc) \Comment{Visit step}
    \State setIsVisited(selectedFunc)
    \State candidates = candidates - selectedFunc
    \State trajs.append([selectedFunc, funcContent]) \Comment{Save trajs}
    \State \textit{// Begin Expand Step}
    \State imports = matchExtract(selectedFunc) \Comment{Get all import statements in the file}
    \State neighbors = $\text{LLM}_{construct}$(imports, funcContent) \Comment{Dynamically expand dependency graph}
    \State candidates = candidates $\cup$ neighbors
    \State \textit{// End Expand Step}
    \State hop += 1
    \EndWhile
    \State contextFunctions = $\text{LLM}_{extract}$(tarFunc, trajs)
    \State \Return contextFunctions
\end{algorithmic}
\label{AL1}
\end{algorithm}

\subsection{Similarity-based Context Retrieval}\label{step2}
This phase takes the signature of the target function and the repository as input, which aims to retrieve a list of similar functions that are useful for generating the target function.
Since existing studies have already demonstrated that function-level context is more effective than code-slice-based context~\cite{AllianceCoder}, we only retrieve similar functions.
As prior studies~\cite{GraphCodeAgent, tao2025retrieval, jiang2025agentic} have shown that dense retrieval outperforms sparse retrieval, we adopt embedding-based dense retrieval in this phase.
Specifically, a static parser is used to extract all functions from the entire repository, embed them using an embedding model, and construct an index using their embeddings (denoted as \textit{function embeddings}).
Then, \appname embeds the signature of the target function using the same embedding model and compute the cosine similarity between the obtained embedding and the function embeddings in the repository.
Finally, \appname selects the top-k most similar functions in descending order of similarity.

\subsection{Generation}\label{step3}
This phase takes the signature of a target function, the related functions acquired by the \retriever, and the similar functions acquired by the similarity-based context retrieval as input, and returns the generated completed target function.
Specifically, to avoid redundant context from interfering with the LLM’s reasoning, the retrieved relevant functions and similar functions are merged into a single deduplicated set.
The functions in this set are regarded as the context of the target function.
Then the LLM is prompted to generate the target function based on its signature and the provided context.
Finally, to prevent the generated code from becoming non-executable due to indentation errors, a post-processing stage is applied to adjust the indentation of the completed function.
The post-processed result is then returned as the final completed implementation of the target function.
The detailed prompt templates of all phases can be found in our \hyperref[rp]{replication package}.

\section{Experiment Setup}

\subsection{Datasets \& Metrics}

To evaluate the performance of \appname on repository-level function generation tasks, we employ two widely used benchmarks, including CoderEval-Python~\cite{CoderEval} and DevEval~\cite{DevEval}, following prior work~\cite{reposcope, cocogen}.
These two benchmarks require LLMs to generate the implementation of a target function in a repository according to its signature, and provide a test suite for each task to assess the correctness of the generated function.

\begin{itemize}[left=0pt, topsep=0em]
\item \textbf{CoderEval-Python}~\cite{CoderEval} is the Python subset of CoderEval containing 230 instances.
It is designed to assess code generation capabilities across six levels of contextual dependency.

\item \textbf{DevEval}~\cite{DevEval} is a manually annotated benchmark designed to evaluate the programming capabilities of large language models within real-world code repositories. 
It contains 1,825 instances extracted from 117 codebases. 
Since DevEval lacks a sandboxed evaluation environment, we find that 209 instances fail in our environment when executing the provided test suites to evaluate the ground-truth solutions.
We provide detailed information about these cases in the \hyperref[rp]{replication package}. 
For consistency, we treat all 209 instances as failures for \appname and all baselines, and compute the final metrics over the full set of 1,825 instances.

\end{itemize}

To evaluate the effectiveness, we adopt the widely used metric \textbf{Pass@1}~\cite{chen2021evaluating, evalplus} following prior work~\cite{reposcope}.
For efficiency evaluation, we record the \textbf{runtime (in seconds)} of each method and count the number of tokens (\textbf{\#Token}) involved in LLM interactions.

\subsection{Baselines}

We mainly compare \appname with open-source state-of-the-art methods that focus on repository-level function generation, including RLCoder~\cite{rlcoder}, AlignCoder~\cite{aligncoder}, and RepoScope~\cite{reposcope}.
We additionally include direct prompting and a simple RAG approach as baselines to highlight the effectiveness of the context collected by \appname.
The baselines are described as follows:

\begin{itemize}[left=0pt, topsep=0em]
\item \textbf{Direct}: It generates the function body directly from the function signature without providing any context.

\item \textbf{SimpleRAG}: It retrieves a set of similar code slices by computing the cosine similarity between the embedding of the signature of the target function and code slices in the repository, and uses them as the context for generation.
Following prior work~\cite{reposcope, rlcoder}, the code slices are obtained by using a sliding window algorithm~\cite{repocoder} to segment the code files in the repository.

\item \textbf{RLCoder} is a reinforcement learning-based framework for repo-level code completion that optimizes retriever behavior through perplexity feedback.

\item \textbf{AlignCoder} constructs an enhanced query by generating multiple candidate completions to bridge the semantic gap between the initial query and the target code, and uses reinforcement learning to train a retriever to learn how to leverage the inference information contained in the enhanced query.

\item \textbf{RepoScope} builds a Repository Structural Semantic Graph (RSSG) and retrieves a four-view context that integrates structural and similarity-based information. 
It further introduces a call-chain prediction method that leverages repository-level structural semantics to identify the target function’s callees more accurately.

\end{itemize}

\subsection{Research Questions}

\begin{itemize}[left=0pt, topsep=0em]
    \item \textbf{RQ1. Effectiveness}: How effective is \appname in repository-level function generation?

    \item \textbf{RQ2. Efficiency}: How efficient is \appname in repository-level function generation?

    \item \textbf{RQ3. Ablation}: How do the key components of \appname contribute to its effectiveness?

    \item \textbf{RQ4. Generalizability}: Can the context collected by \retriever be effectively combined with different similarity-based methods?
\end{itemize}

\subsection{Implementation Details}

We conduct experiments with Qwen3-Coder~\cite{qwen3}, DeepSeek-v3.2~\cite{deepseekv3.2}, and GPT-4o-mini~\cite{hurst2024gpt}.
They are from different families and cover both open-source and closed-source models.
For Qwen3-Coder, we employ Qwen3-Coder-30B-Instruct and deploy it locally via vLLM~\cite{vllm} on 2 Nvidia A100 GPUs.
For DeepSeek-v3.2 and GPT-4o-mini, we access them through their official API services.

In the official implementation released by RepoScope~\cite{reposcope}, the corresponding test suites of the target function are not masked from the repository, which may lead to artificially inflated performance. 
To prevent potential data leakage, we mask both the signature and implementation of the target function in the repository and provide only the signature to the LLM as the task specification in the prompt. 
Furthermore, all methods are restricted from accessing non-code files or test files.

To ensure a fair comparison, we use the Qwen3-Embedding-0.6B model~\cite{qwen3-embedding} via vLLM on 1 Nvidia A100 GPU, as the embedding model for embedding-based similarity calculation in all the methods except RLCoder and AlignCoder, since they train their own embedding models. 
We reuse the embedding models released by RLCoder~\cite{rlcoder} and AlignCoder~\cite{aligncoder}.
Since these two methods are originally designed for repository-level code completion, we adapt them to the CoderEval~\cite{CoderEval} and DevEval~\cite{DevEval} benchmarks by replacing their prompt templates with those of \appname.

For \appname, we set the maximum hop of the multi-hop reasoning to 10, since preliminary experiments setting it in [1,20] show that when the maximum hop exceeds 10, \appname’s performance begins to plateau, while the token consumption increases by 19.1\%.
The number of retrieved code snippets of all methods is controlled consistently with RepoScope~\cite{reposcope}, i.e., providing the Top-10 similar functions for the similarity-based context, and providing the Top-5 related functions for extra context.

When interacting with LLMs, we employ greedy sampling (temperature = 0) to ensure the reproducibility of the LLM’s sampling results.
For measuring time cost, we record the execution time of all the methods on Intel Xeon Platinum 8358P CPUs, using a serial execution paradigm.

\section{Experiment Results}

\begin{table}[t!]
\centering
\caption{Performance comparison of different methods across different LLMs and benchmarks.}
\setlength{\tabcolsep}{3pt}
\begin{adjustbox}{width=\columnwidth}
\begin{tabular}{clcccc}
\toprule
\multirow{2}{*}{\textbf{}} & \multirow{2}{*}{\textbf{Method}} & \multicolumn{3}{c}{\textbf{Model}} &\multirow{2}{*}{\textbf{Avg.}}\\
\cmidrule{3-5}
 & & \textbf{Qwen3-Coder-30B}& \textbf{DeepSeek-v3.2}& \textbf{GPT-4o-mini} &\\
\midrule
\multirow{6}{*}{\rotatebox{90}{\textbf{CoderEval}}}
&Direct& 24.35 \up{103.2}& 24.00 \up{112.7}&  20.78 \up{94.2} &23.04 \up{103.83}\\
 &SimpleRAG& 36.17 \up{36.8}& 39.83 \up{28.1}&  28.26 \up{42.8} &34.75 \up{35.14}\\
 &RLCoder& 35.65 \up{38.8}& 42.17 \up{21.0}&   34.78 \up{16.0} &37.52 \up{25.16}\\
 & AlignCoder& 38.17 \up{29.6}&  38.26 \up{33.4}&  34.78 \up{16.0} &37.07 \up{26.68}\\
 & RepoScope& 42.96 \up{15.2}&  43.22 \up{18.1}&  36.00 \up{12.1} &40.72 \up{15.32}\\
\rowcolor{highlight} & \appname&  \textbf{49.48} &  \textbf{51.04}& \textbf{40.35} &\textbf{46.96}\\
\midrule
 \multirow{6}{*}{\rotatebox{90}{\textbf{DevEval}}}
 &Direct&  12.90 \up{263.4}&  13.91 \up{239.7}&  5.18 \up{448.1} &10.66 \up{283.11}\\
 & SimpleRAG& 27.34 \up{71.5}& 31.87 \up{48.3}& 15.37 \up{84.7} &24.86 \up{64.28}\\
 & RLCoder& 23.97 \up{95.6}& 27.45 \up{72.1}& 15.90 \up{78.6} &22.44 \up{82.00}\\
 & AlignCoder& 23.60 \up{98.6}& 27.70 \up{70.6}& 16.46 \up{72.5} &22.58 \up{80.87}\\
 & RepoScope& 43.06 \up{8.9}& 45.39 \up{4.1}& 25.54 \up{11.2} &38.00 \up{7.47}\\
\rowcolor{highlight} & \appname& \textbf{46.88}& \textbf{47.25}& \textbf{28.39} &\textbf{40.84}\\
\bottomrule

\end{tabular}
\end{adjustbox}
\label{rq1}
\end{table}

\subsection{RQ1: Effectiveness}

To validate the effectiveness of \appname in repository-level code generation, we conduct experiments with three different LLMs and compare \appname with the baselines. 
Table~\ref{rq1} presents the experimental results on CoderEval and DevEval, respectively.
Experimental results show that on CoderEval and DevEval, \appname achieves an improvement of 12.1\%-42.8\% and 4.1\%-98.6\%, respectively, compared to the RAG-based baselines, which are significantly more effective in repository-level code generation than the baselines.
This demonstrates that by simulating human developers' behavior of collecting contextual information, \appname can effectively retrieve useful context to support repository-level code generation.
Moreover, compared with \textit{similarity-based methods}, i.e., SimpleRAG, RLCoder, and AlignCoder, \appname achieves an average improvement of 29.13\% and 76.94\% on CoderEval and DevEval, respectively. 
Compared with the \textit{graph-based method}, RepoScope, \appname still achieves an improvement of 4.1\%--18.1\%, indicating that, unlike rule-based retrieval algorithms, leveraging LLM to assess the usefulness of functions enables \appname to more flexibly retrieve truly helpful context.
Despite this, \appname is also 7$\times$ faster than RepoScope, making it an accurate and efficient approach.

The improvement in DevEval is greater than in CoderEval because, compared to CoderEval, the repositories in DevEval are larger and contain more complex dependency relations~\cite{DevEval}, making it more difficult for similarity-based methods to retrieve useful context.
This indicates that compared with similarity-based methods, \appname can effectively acquire useful context from complex dependency relationships through multi-hop reasoning.

\begin{figure}[t]
    \centering
    \includegraphics[width=\linewidth]{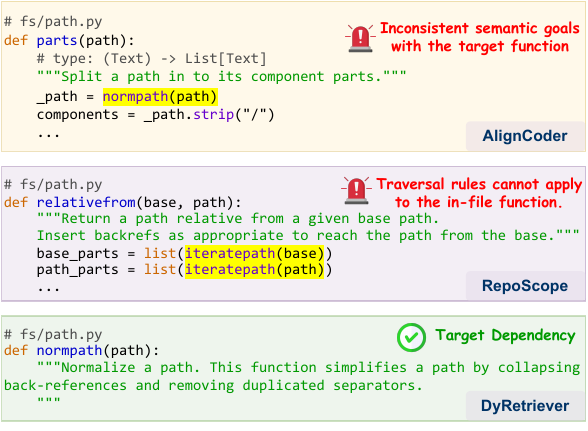}
    \caption{An example of the retrieved context and generated code by \appname and baselines. The target function is the same as the one in Figure~\ref{fig:motivation}.}
    \label{fig:result_case}
\end{figure}

To further investigate why \appname outperforms other baselines, we manually inspect the experimental results and summarize two main reasons.
First, by simulating the behaviors of human developers, \retriever provides additional useful context about the target function compared with the similarity-based baseline.
For example, Figure~\ref{fig:result_case} presents the context retrieved and the code generated by \appname and two best-performing baseline methods for the example given in Figure~\ref{fig:motivation}.
AlignCoder~\cite{aligncoder} retrieves the similar function \texttt{parts}. 
However, the semantic goals of the two functions are inconsistent: \texttt{parts} encodes path representations, while \texttt{iteratepath} requires pure path components. 
Reusing \texttt{parts} results in incorrect behavior in \texttt{iteratepath}, which indicates that similar functions are not necessarily relevant to the requirement and may even mislead the generation.
In contrast, \appname retrieves the full implementation of \texttt{normpath}, enabling the LLM to obtain the complete context and examples necessary to generate the target function, leading to a correct solution.

Second, \appname can more flexibly obtain useful context related to the target function through dependency graphs. 
For example, RepoScope’s call chain prediction algorithm is manually designed, which relies on code entities imported by the target functions as entry points to start traversing the code graph.
However, the target function \texttt{iteratepath} and its dependency function \texttt{normpath} are in the same code file, leading to the lack of import statements.
Consequently, RepoScope cannot establish an effective traversal entry point for this case and thus fails to retrieve \texttt{normpath}.
Although the call chain predicted by RepoScope includes the caller function \texttt{relativefrom} of \texttt{iteratepath}, this function does not provide enough information about \texttt{normpath}, which leads the LLM to use an incorrect method for the path normalization operation during generation.
This rule-free approach, which relies on an LLM to assess helpfulness, enables flexible and accurate multi-hop reasoning, allowing \appname to obtain the desired context.

\answerRQ{
Compared to RAG-based baselines, \appname achieves an average improvement of 25.63\% on CoderEval and 59.73\% on DevEval in Pass@1, demonstrating its effectiveness in performing retrieval by simulating the behavior of humans.
}

\subsection{RQ2: Efficiency}

\begin{table}[t!]
\centering
\caption{Efficiency analysis on CoderEval using Qwen3-Coder.}
\begin{threeparttable}
\begin{adjustbox}{width=\columnwidth}
\setlength{\tabcolsep}{3pt}
\begin{tabular}{lcccccccc}
\toprule
 \multirow{2}{*}{\textbf{Method}} & \multicolumn{5}{c}{\textbf{Time Cost}}& \multicolumn{2}{c}{\textbf{Inference Cost}} &\multirow{2}{*}{\textbf{Train?}}\\
\cmidrule(lr){2-6}\cmidrule(lr){7-8}
 & \textbf{Indexing}&\textbf{Retrieval}& \textbf{Generation}& \textbf{Total}  & \textbf{imp.}& \textbf{\#Token}&\textbf{\$Cost} &\\
\midrule
Direct& 0&0& 9.65& 9.65& \nofast{0.07}& 3.29k& 0.0016&\xmark\\
SimpleRAG& 100.87&17.36&  9.27& 127.50& \nofast{0.86}& 3.65k& 0.0016&\xmark\\
RLCoder& 103.09&45.23&  4.71& 153.03& \fast{1.03}& 2.81k& 0.0010&\cmark\\
AlignCoder& 100.89&49.09&  4.25& 154.23& \fast{1.04}& 3.65k& 0.0013&\cmark\\
RepoScope& 875.32&209.72&  11.19& 1096.23& \fast{7.40}& 5.11k& 0.0020&\xmark\\
\rowcolor{highlight}
\appname &
85.62 &
41.09/50.01\tnote{1} &
12.46 &
148.09  & \fast{1.00}&
13.09k &
0.0039 &
\xmark \\
\bottomrule
\end{tabular}
\end{adjustbox}
\begin{tablenotes}
\footnotesize
\item[1] 41.09s for \retriever and 50.01s for similarity-based retrieval.
\end{tablenotes}
\end{threeparttable}
\label{rq3}
\end{table}

To evaluate the efficiency of \appname in repository-level code generation tasks, we compare its cost with that of other baselines from two perspectives: time cost and inference cost.
For time cost, we measure the execution time of the indexing, retrieval, and generation phases for RAG-based methods. 
Since \retriever and the similarity-based context retrieval phase of \appname can execute in parallel, we include the maximum of the two as part of the total time. 
For inference cost, we obtain the token usage information and compute the dollar cost according to the official pricing.
We conduct measurements on CoderEval using a locally deployed Qwen3-Coder model (with a generation speed of about 400 tokens/s).

As shown in Table~\ref{rq3}, \appname incurs a lower total time cost than all baselines except Direct and SimpleRAG, and is 7.4 times faster than the graph-based RepoScope.
In the \textit{indexing phase}, \appname is faster than all RAG-based baselines. 
This is because, compared with similarity-based baselines, \appname only extracts the functions in the repository via static analysis, avoiding finer-grained slicing and embedding. 
Compared with graph-based baselines, \appname dynamically constructs a partial dependency graph instead of constructing a full repository-level code graph, reducing the indexing time overhead by roughly an order of magnitude.
Moreover, \appname constructs partial dependency graphs using an LLM and discards them after use, thereby avoiding the computational overhead incurred by static-analysis-based graph construction. 

In the \textit{retrieval phase}, \appname incurs a time cost comparable to that of similarity-based baselines.
The time cost of \retriever is slightly higher than that of similarity-based retrieval, as partial graph construction requires multiple rounds of LLM reasoning. 
Nevertheless, as graph-based baselines (i.e., RepoScope) rely on manually designed, high-complexity algorithms for call chain prediction, \appname remains nearly four times faster in the retrieval phase.

In the \textit{generation phase}, the time cost depends on the length of the retrieved context. 
Since both \appname and RepoScope incorporate context from two perspectives, their generation time is slightly longer than that of the other baselines.
Overall, \appname achieves time efficiency comparable to similarity-based baselines while substantially outperforming graph-based baselines.

This efficiency gain is partly attributable to constructing only partial dependency graphs, which reduces parsing overhead, and partly to relying on LLMs for dependency graph construction, which avoids the maintenance cost incurred when the codebase changes.
In real-world development scenarios, developers are typically not limited to generating a single function, which causes dependencies in certain parts of the repository to change frequently. 
Under such conditions, constructing and maintaining a global graph index with static tools requires repeatedly rebuilding parts of~\cite{samhi2025you, ryder2001change}, or even the entire~\cite{reps1995precise} graph index when dependencies evolve. 
Since building such a graph index is a highly time-consuming operation, this process slows down the developers’ workflow. 

Table~\ref{rq3} also shows that \appname consumes 1.95-2.43 times more tokens. We argue that this is often acceptable for developers for two reasons.
First, the absolute dollar cost remains low, as generating each function costs only \$0.0039 on average.
Second, \appname achieves superior effectiveness while maintaining a high generation speed.
With a relatively low absolute dollar cost, \appname outperforms graph-based baselines while operating at a speed close to that of similarity-based baselines.
Finally, as an out-of-the-box method, \appname requires no training, which aligns well with developer practices in real application scenarios~\cite{pan2025measuring}.

\answerRQ{
\appname achieves a 7.4$\times$ speedup over the graph-based baseline, performs almost on par with similarity-based baselines, and incurs an average cost of only \$0.0039 per generated function, demonstrating its efficiency.
}

\subsection{RQ3: Ablation}

\begin{table}[t!]
\centering
\caption{Ablation results with two representative models.}
\begin{adjustbox}{width=\columnwidth}
\setlength{\tabcolsep}{3pt}
\begin{tabular}{clcccc}
\toprule
\multirow{2}{*}{} & \multirow{2}{*}{\textbf{Method}} & \multicolumn{2}{c}{\textbf{Qwen3-Coder-30B}} & \multicolumn{2}{c}{\textbf{GPT-4o-mini}} \\
\cmidrule(lr){3-4}\cmidrule(lr){5-6}
& & \textbf{Pass@1} & \textbf{\#Token} & \textbf{Pass@1} & \textbf{\#Token} \\
\midrule

\multirow{4}{*}{\rotatebox{90}{\textbf{CoderEval}}}
& \cellcolor{highlight}\appname
& \cellcolor{highlight}49.48
& \cellcolor{highlight}13.09k
& \cellcolor{highlight}40.35
& \cellcolor{highlight}11.81k \\
& \hspace{1em} w/o Similarity-based & 42.87\down{13.4} & 11.61k\down{11.3} & 33.83\down{16.2} & 10.13k\down{14.2} \\
& \hspace{1em} w/o \retriever & 41.57\down{16.0} & 4.67k\down{64.3} & 34.00\down{15.7} & 4.60k\down{61.0} \\
& \hspace{1em} w/o Multi-hop Reasoning & 46.52\down{6.0} & 6.81k\down{47.9} & 37.48\down{7.1} & 6.69k\down{43.4} \\

\midrule

\multirow{4}{*}{\rotatebox{90}{\textbf{DevEval}}}
& \cellcolor{highlight}\appname
& \cellcolor{highlight}46.88& \cellcolor{highlight}112.41k& \cellcolor{highlight}28.39& \cellcolor{highlight}102.38k\\
& \hspace{1em} w/o Similarity-based & 35.71\down{23.8}& 100.09k\down{11.0}& 26.56\down{6.5}& 88.11k\down{13.9}\\
& \hspace{1em} w/o \retriever & 22.94\down{51.1}& 45.54k\down{59.5}& 15.55\down{45.2}& 37.38k\down{63.5}\\
& \hspace{1em} w/o Multi-hop Reasoning & 42.40\down{9.6}& 69.31k\down{38.3}& 25.76\down{9.3}& 63.49k\down{38.0}\\
\bottomrule
\end{tabular}
\end{adjustbox}
\label{rq2}
\end{table}

\appname consists of two key context components, i.e., similarity-based context retrieval and \retriever.
Within \retriever, multi-hop reasoning with partial graphs serves as a core subcomponent.
To analyze their contributions to \appname's effectiveness, we first separately eliminate each of the key context components to construct two main variants (\textit{w/o Similarity-based} and \textit{w/o \retriever}).
Then, we further construct a variant that removes only the multi-hop reasoning with partial graphs (\textit{w/o Multi-hop Reasoning}).
This variant directly uses the entry points obtained in Section~\ref{step1.1} as the related functions retrieved by \retriever to generate the target function.
We evaluate all variants on CoderEval and DevEval using the open-source model Qwen3-Coder-30B and the closed-source model GPT-4o-mini, as \appname achieves better performance gains on CoderEval across both models.

As shown in Table~\ref{rq2}, \appname outperforms all three variants in Pass@1 across CoderEval and DevEval.
Specifically, removing the similarity-based context retrieval leads to noticeable performance drops on both datasets, with decreases of 13.4\% and 16.2\% on CoderEval, and 23.8\% and 6.5\% on DevEval under Qwen3-Coder-30B and GPT-4o-mini, respectively.
Similarly, removing \retriever results in even larger degradations, reaching up to 16.0\% and 51.1\% on CoderEval and DevEval, respectively.
These results demonstrate that both types of context used in \appname contribute to generating the target function, and that the useful information they provide exhibits a certain degree of complementarity.

In addition, we find that removing the multi-hop reasoning with partial graphs leads to consistent performance drops across both datasets (6.0\%--7.1\% on CoderEval and 9.3\%--9.6\% on DevEval).
This component simulates how human developers inspect code along the dependency graph. Without it, \appname becomes less capable of identifying helpful functions that can help to generate the target function, since multi-hop reasoning enables the LLM to flexibly traverse the code dependency graph and assess whether a code snippet is helpful to the target function.
Although removing multi-hop reasoning leads to a certain performance degradation for \appname, its Pass@1 still outperforms the variant (i.e., \textit{w/o \retriever}) that relies solely on similarity-based context.
This is because the \textit{w/o Multi-hop Reasoning} variant still preserves the simulation of how human developers select entry points based on their understanding of the repository.
Because this simulation provides code snippets as context from a perspective different from similarity, it can collect additional contextual information that is useful for generating the target function. 
These results further support the effectiveness of \appname in retrieving context by simulating how human developers search for relevant information.

\answerRQ{The similarity-based context retrieval and \retriever exhibit strong complementarity, and removing either of them leads to a 6.5\%-23.8\% and 15.7\%-51.1\% performance drop, respectively. 
Multi-hop reasoning with partial graphs plays a crucial role in \retriever, which contributes 6.0\%-9.6\% of the overall performance.}

\begin{table}[t!]
\centering
\caption{Performance of different similarity-based method combined with \appname on CoderEval.}
\begin{adjustbox}{width=0.8\columnwidth}
\setlength{\tabcolsep}{3pt}
\begin{tabular}{lcc}
\toprule
 \multirow{2}{*}{\textbf{Method}} & \multicolumn{2}{c}{\textbf{Model}}\\
\cmidrule(lr){2-3}
  & \textbf{Qwen3-Coder-30B}&\textbf{GPT-4o-mini}\\
\midrule
BM25& 41.71&34.87\\
\rowcolor{highlight}  \hspace{1em} + \retriever& 48.70\up{14.65}&37.48\up{7.50}\\
RLCoder& 35.65&34.78\\
\rowcolor{highlight}  \hspace{1em} + \retriever& 43.91\up{23.17}&37.91\up{9.00}\\
AlignCoder& 38.17&34.78\\
\rowcolor{highlight}  \hspace{1em} + \retriever& 50.00\up{30.99}&37.57\up{8.00}\\

\bottomrule

\end{tabular}
\end{adjustbox}
\label{rq4}
\end{table}

\subsection{RQ4: Generalizability}

To investigate whether the context collected by \retriever can generalize and be combined with various similarity-based baselines, we conduct experiments using the same models on CoderEval. 
We replace the similarity-based context retrieval in \appname with the context retrieved by RLCoder and AlignCoder to examine \appname’s generalizability to dense retrieval. 
Additionally, we incorporate context retrieved using BM25 to evaluate \appname’s generalizability to sparse retrieval, since BM25 is one of the most popular sparse retrieval algorithms~\cite{lu2022reacc} and has already demonstrated strong performance in the code search domain~\cite{ma2025alibaba}. 
We do not combine \retriever with RepoScope, as RepoScope already incorporates graph-based context.

Table~\ref{rq4} presents the Pass@1 scores of the two models on CoderEval.
For Qwen3-Coder-30B, combining similarity-based baselines with \retriever yields an improvement of 14.65\%-30.99\%. 
For GPT-4o-mini, the improvement ranges from 7.5\% to 9.0\%.
These results indicate that the context collected by \retriever exhibits low coupling with the context retrieved by both types of similarity-based methods and shows strong complementarity with them.
This benefit arises from \retriever’s ability to retrieve context useful for generating the target function by simulating how human developers retrieve information. 
Through performing multi-hop reasoning along the dependency graph, \retriever explicitly examines each function along the traversal path to determine its usefulness. 
As this perspective differs from similarity-based retrieval, the two retrieval methods are complementary.
Notably, the improvement of \retriever on GPT-4o-mini is smaller than that on Qwen3-Coder-30B.
This may be due to inherent differences in the reasoning capabilities of different models, and also suggests that \retriever exhibits stronger effectiveness when applied to models with higher reasoning capacity.

\answerRQ{
\retriever can be integrated with existing similarity-based methods to achieve 7.5\%-30.99\% performance improvement. This not only demonstrates the generalizability of \retriever but also highlights its advantages and optimization potential as a plug-and-play module.
}

\section{Discussion}

\subsection{Are Graphs Constructed by LLMs Accurate?}

Constructing a dependency graph for a code repository is far from trivial, especially for dynamically typed languages such as Python~\cite{bouzenia2024dypybench}.
Prior work~\cite{bouzenia2024dypybench, salis2021pycg} has shown that existing static analysis tools cannot guarantee sufficient accuracy, and an inaccurate dependency graph may limit the upper bound of \retriever’s performance. 
Since \retriever relies on the LLM to identify dependency relations between functions, and LLM hallucination is unavoidable, it is necessary to ensure the quality of the dependency graph constructed by the LLM.
To investigate whether the dependency relations identified by the LLM are sufficiently accurate for \retriever after the post-processing, we manually inspected the construction process in 60 \textit{expand} operations and counted the number of cases in which the constructed dependencies are completely accurate.
Manual inspection shows that for \textbf{93.22\%} of the functions, the LLM can accurately identify all of their dependencies and construct the dependency graph.

Moreover, we also examine why the LLM is able to accurately identify dependency relations and summarize two main reasons.
First, \retriever requires the LLM to identify only downstream dependencies (i.e., the callee functions). 
Unlike identifying upstream dependencies (i.e., the caller functions), which requires a global analysis, identifying downstream dependencies only requires examining the callees of a given function. 
Since recognizing callees is far easier for an LLM than tracing callers~\cite{venkatesh2025empirical}, the LLM can construct the dependency graph with relatively high accuracy.
Second, identifying the dependencies of an individual function is relatively easy, since the dependencies of a single function are few and fairly explicit. 
For example, among the 60 cases we examined, only one function had more than ten dependencies, and about 40\% of the functions had fewer than five dependencies. 
Moreover, their dependencies can be found in either the same code file or the import statements of the file.
Thus, the LLM can easily infer its origins from the import statements and thereby identify the dependencies.
Despite this, the LLM still cannot achieve 100\% identification accuracy, which may occur when the given function is long (spanning hundreds of lines), causing redundant context to interfere with the LLM’s recognition, or when the function has many dependencies (above 10), leading the LLM to identify only a subset of them.
But this accuracy is acceptable, since it is already comparable to the accuracy (77.4\%-99.2\%) achieved by static analysis tools~\cite{salis2021pycg}.

\subsection{Threats to Validity}
\subsubsection{Internal Validity}
\textbf{Hyperparameter settings.}
The number of retrieved code snippets affects the effectiveness of \appname compared with the baselines. To ensure a fair comparison, we align the number of the two types of context retrieved by \appname with that retrieved by RepoScope~\cite{reposcope}.
The maximum number of hops in the multi-hop reasoning may affect the effectiveness of \appname.
To mitigate this threat, we conduct a preliminary exploration on CoderEval~\cite{CoderEval} with hop counts ranging from 0 to 20, and select the best-performing setting.

\textbf{Baseline implementation.} 
Since AlignCoder~\cite{aligncoder} and RLCoder~\cite{rlcoder} are originally proposed for repository-level code completion, we adapt them to the code generation setting for comparison by modifying their prompt template. 
To mitigate performance variations introduced by this adaptation, we reuse all released artifacts from the original papers, including their embedding models, workflows, and related components.
We follow all hyperparameter settings released in the original papers to ensure the consistency of reproduced results, to mitigate such threats.

\subsubsection{External Validity}
\textbf{Generalization to other programming languages.}
We currently evaluate our method only on Python, and the generalization to other programming languages may not yet be fully validated.
However, Python is one of the most popular and widely used languages in repository-level code generation benchmarks~\cite{CoderEval, DevEval, crosscodeeval, evocodebench, repocoder}, which can demonstrate the effectiveness of \appname.
Moreover, as a dynamically typed language, it is more challenging to identify dependencies in the application scenarios. 
Statically typed languages such as Java provide clearer and more easily extractable type relations, which may make dependency identification easier for LLMs and thus mitigate this threat to some extent.
Since \appname does not rely on static analysis to obtain effective information, it can be readily generalized to other programming languages.

\section{Related Work}

In recent years, repository-level code generation has attracted extensive attention from researchers, as it more closely reflects real-world production scenarios.
Unlike simple code generation tasks, which only need to generate single functions~\cite{evalplus, fan2025fait, fan2026recode}, target functions of repository-level code generation typically exhibit complex contextual dependencies.
A lot of benchmarks have been proposed to evaluate the capability of LLMs on repository-level code generation tasks, such as RepoEval~\cite{repocoder}, CoderEval~\cite{CoderEval}, and SWE-bench~\cite{SWE-bench}.
Researchers have introduced retrieval-augmented generation (RAG) techniques into code generation, aiming to provide effective context for repository-level code generation. 
Specifically, RAG methods first build an index for the code repository, then retrieve relevant information based on the requirement description, and finally use the retrieved content as context to prompt the LLM for generation.

Existing repository-level retrieval-augmented code generation methods can generally be categorized into \textit{similarity-based}, \textit{graph-based}, and \textit{agent-based} methods.
\textit{Similarity-based} methods~\cite{aligncoder, repocoder} compute similarity scores using sparse~\cite{BM25} or dense retrieval algorithms~\cite{qwen3-embedding} and select the most similar code snippets as context.
For example, AlignCoder~\cite{aligncoder} first performs sparse retrieval using the BM25~\cite{BM25} algorithm to generate a code draft, and then employs reinforcement learning to train the embedding model AlignRetriever for dense retrieval.
RepoCoder~\cite{repocoder} designs an iterative retrieval mechanism that uses the partially generated code to retrieve similar code snippets from the repository, and then leverages the retrieved results as domain knowledge to assist the model in regenerating the code.
These methods rely solely on similarity and overlook the program-structural semantics in code repositories, which may lead to inferior performance of code generation.

\textit{Graph-based} methods~\cite{jiang2025issue, graphcoder, draco, reposcope} model code dependency relations by static analysis and design retrieval algorithms based on manually designed rules to retrieve code snippets over the code graph, thereby supplementing context from a structured perspective of the code.
For example, GraphCoder~\cite{graphcoder} constructs a code context graph (CCG) to model control flow and data dependence at the line level and manually designs a similarity metric between slices of CCG to perform retrieval and capture context in a coarse-to-fine manner. 
DRACO~\cite{draco} constructs a data-flow graph for the repository and retrieves code snippets by computing similarity between module names through string matching based on import statements, using depth-first search (DFS). 
RepoScope~\cite{reposcope} builds a repository structural semantic graph and designs a call chain prediction algorithm that starts from entities imported by the target function and assigns similarity scores to call chains via DFS to enable multi-view context retrieval.
However, the retrieval algorithms used by these methods~\cite{graphcoder, draco, reposcope} depend on manually designed rules, such as differential similarity metrics, which make them exhibit limited flexibility in real application scenarios. 
In addition, these methods construct complex code dependency graphs through static analysis, which is both time-consuming and difficult to maintain for large repositories with frequently changing dependencies.
Recently, agents have attracted increasing attention in the field of code generation due to their strong reasoning and tool-calling capabilities. 

\textit{Agent-based} methods~\cite{codeagent, GraphCodeAgent, jiang2025issue} design a series of tools that enable agents to access the code repository and retrieve context.
For example, CodeAgent~\cite{codeagent} integrates five programming tools and implements four agent strategies, enabling large language models to autonomously determine retrieval targets.
GraphCodeAgent~\cite{GraphCodeAgent} enables the LLM to identify retrieval targets by establishing a mapping between the requirement graph and the structural-semantic code graph.
However, the retrieval tools equipped by these agents still fall under \textit{similarity-based} or \textit{graph-based} retrieval.
Since the limitations of these two categories of methods remain unresolved, \textit{agent-based} methods still exhibit limited performance.

Different from existing methods, we design \retriever by simulating human developers' behaviors when collecting helpful context.
Specifically, \retriever first uses an LLM to select some entry point functions based on the structure of the repository, and then designs a multi-hop reasoning algorithm to flexibly traverse the code dependency graph and determine whether a function can help to generate the target function.
This breaks the dependence of manually designed rules and makes \retriever adaptively adjust its search direction, resulting in better flexibility.
When constructing code dependency graphs, \retriever uses a partial graph instead of a global graph to reduce the time overhead.
Since no tools can be used to construct a partial graph dynamically, we design an LLM-based partial dependency graph construction method to achieve it.
The partial dependency graph is discarded after use, requiring no additional maintenance cost and allowing the method to adapt to continuously evolving code repositories.

\section{Conclusion and Future Work}

We proposed \retriever, an effective and efficient context retrieval method via partial dependency graphs for repository-level code generation.
By simulating human developers' behavior when collecting helpful context, \retriever leverages multi-hop reasoning over the code dependency graph to collect helpful context flexibly, breaking the dependence of manually designed rules in retrieval algorithms.
In addition, \retriever employs an LLM-driven code dependency graph construction method that dynamically builds partial code dependency graphs during multi-hop reasoning and discards them after use to reduce the unnecessary time overhead of pre-constructing a global dependency graph using static analysis tools, which requires no maintenance.
We integrate \retriever with similarity-based retrieval and propose \appname.
Extensive experiments on the widely used repository-level code generation benchmarks CoderEval and DevEval show that \appname outperforms baseline methods in both generation success rate and speed, achieving up to a 7.4$\times$ speedup and a 98.6\% improvement in Pass@1.

In the future, we plan to extend \appname to more programming languages and evaluate it in more complex code generation scenarios (e.g., feature addition).
We also intend to further investigate LLM-based code dependency graph construction techniques to achieve more accurate dependency graph generation. 
In addition, we will explore leveraging the plug-and-play nature of \appname to strengthen agent-based code generation methods.

\section*{Data Availability}
\phantomsection
\label{rp}
The replication package is available at \url{https://doi.org/10.5281/zenodo.19235859}.
Our official repository is available at \url{https://github.com/ZJU-CTAG/DyCoder}.

\begin{acks}
This research is supported by the National Natural Science Foundation of China (No. 92582107 and No. 62302430) and Zhejiang Provincial Natural Science Foundation of China (No. LZ25F020003 and No. LQ24F020017).
\end{acks}

\balance
\bibliographystyle{ACM-Reference-Format}
\bibliography{references}

\end{document}